\documentclass[prl,reprint,superscriptaddress,preprintnumbers]{revtex4-1}
\usepackage{amsfonts, amssymb, amsmath}
\usepackage{mathrsfs}
\usepackage{hyperref}

\newcommand{\dvol}{d\mathrm{vol}}

\newcommand{\Vol}{\mathrm{Vol}}

\newcommand{\ib}{{\bar \imath}}

\newcommand{\parfrac}[2]{\frac{\partial #1}{\partial #2}}

\newcommand{\wb}{\overline}

\newcommand{\ie}{\textit{i.e.}}

\newcommand{\mat}[1]{\begin{pmatrix} #1 \end{pmatrix}}

\newcommand{\be}{\begin{equation}} \newcommand{\ee}{\end{equation}}
\newcommand{\bea}{\begin{equation} \begin{aligned}} \newcommand{\eea}{\end{aligned} \end{equation}}

\newcommand{\cF}{\mathcal{F}}
\newcommand{\cG}{\mathcal{G}}

\newcommand{\cI}{\mathcal{I}}

\newcommand{\cK}{\mathcal{K}}
\newcommand{\cL}{\mathcal{L}}
\newcommand{\cM}{\mathcal{M}}
\newcommand{\cN}{\mathcal{N}}
\newcommand{\cO}{\mathcal{O}}

\newcommand{\cQ}{\mathcal{Q}}

\newcommand{\cV}{\mathcal{V}}

\newcommand{\cZ}{\mathcal{Z}}

\newcommand{\bR}{\mathbb{R}}
\newcommand{\bZ}{\mathbb{Z}}
\newcommand{\fg}{\mathfrak{g}}

\newcommand{\fp}{\mathfrak{p}}
\newcommand{\fq}{\mathfrak{q}}

\newcommand{\su}{\mathfrak{su}}

\DeclareMathOperator{\Tr}{Tr}

\DeclareMathOperator{\re}{\mathbb{R}e}
\DeclareMathOperator{\im}{\mathbb{I}m}

\begin{document}

\title{Exact microstate counting for dyonic black holes in AdS$\boldsymbol{_4}$}

\date{\today}

\author{Francesco Benini}

\affiliation{SISSA, via Bonomea 265, 34136 Trieste, Italy --- INFN, Sezione di Trieste}
\affiliation{Blackett Laboratory, Imperial College London, London SW7 2AZ, United Kingdom}

\author{Kiril Hristov}

\affiliation{INRNE, Bulgarian Academy of Sciences, Tsarigradsko Chaussee 72, 1784 Sofia, Bulgaria}

\author{Alberto Zaffaroni}

\affiliation{Dipartimento di Fisica, Universit\`a di Milano-Bicocca, I-20126 Milano, Italy}
\affiliation{INFN, sezione di Milano-Bicocca, I-20126 Milano, Italy}

%\emailAdd{f.benini@imperial.ac.uk}
%\emailAdd{khristov@inrne.bas.bg}
%\emailAdd{alberto.zaffaroni@mib.infn.it}

\begin{abstract}

We present a counting of microstates of a class of dyonic BPS black holes in AdS$_4$ which precisely reproduces their Bekenstein-Hawking entropy. The counting is performed in the dual boundary description, that provides a non-perturbative definition of quantum gravity, in terms of a twisted and mass-deformed ABJM theory. We evaluate its twisted index and propose an extremization principle to extract the entropy, which reproduces the attractor mechanism in gauged supergravity.

\end{abstract}


\preprint{SISSA 41/2016/FISI}

\maketitle

Supersymmetric black holes in string theory constitute important models to test fundamental questions about quantum gravity in a relatively simple setting. The main question we would like to address here is the origin of the black hole (BH) entropy, which statistically is expected to count the number of degenerate BH configurations. String theory provides a microscopic explanation for the entropy of a class of asymptotically flat black holes \cite{Strominger:1996sh}. Much less is known about asymptotically AdS ones in four or more dimensions.

In principle AdS/CFT \cite{Maldacena:1997re} provides a non-perturbative definition of quantum gravity in asymptotically AdS space, as a dual boundary quantum field theory (QFT). The BH microstates appear as particular states in the boundary description. The difficulty with this approach is the need to perform computations in a strongly coupled QFT, but the development of exact non-perturbative techniques makes progress possible.
We recently reported \cite{Benini:2015eyy} on a particular example of magnetically charged BPS black holes in AdS$_4$ \cite{Cacciatori:2009iz} with a known field theory dual---topologically twisted ABJM theory. Using the technique of supersymmetric localization we were able to calculate in an independent way the (regularized) number of ground states of the theory and successfully match it with the leading macroscopic entropy of the black holes.

In this Letter we discuss a function $Z(u_a)$ that encodes the quantum entropies of static dyonic BPS black holes in AdS$_4$, computed non-perturbatively from the dual QFT description, and show that its leading behavior reproduces the Bekenstein-Hawking entropy \cite{Bekenstein:1973ur,*Hawking:1974sw}.

In particular we show that, at leading order, the entropy of BPS black holes with magnetic charges $\fp_a$, electric charges $\fq_a$ and asymptotic to AdS$_4\times S^7$ can be obtained by extremizing the quantity
\be
\cI = \log Z(u_a) - i \sum\nolimits_a u_a \fq_a
\ee
with respect to a set of complexified chemical potentials $u_a$ for the global $U(1)$ flavor symmetries of the boundary theory. $Z(u_a)$ is the topologically twisted index \cite{Benini:2015noa} of the ABJM theory \cite{Aharony:2008ug} which explicitly depends on the magnetic charges $\fp_a$ (see \cite{Hosseini:2016tor,*Hosseini:2016ume} for other examples). The entropy is given by  $S = \cI(\hat u)$ evaluated at the extremum, with a constraint on the charges that $S$ be real positive.

As we will see, the extremization of  $\cI$ is equivalent to the attractor mechanism for AdS$_4$ black holes in gauged supergravity. We also argue, generalizing \cite{Benini:2015eyy}, that the extremization of $\cI$ selects the exact R-symmetry of the superconformal quantum mechanics dual to the AdS$_2$ horizon region. We notice strong similarities between our formalism and those based on Sen's entropy functional \cite{Sen:2008yk} and the OSV conjecture \cite{Ooguri:2004zv}.

{\it The Black Holes}.---%
We consider dyonic BPS BHs that can be embedded in M-theory and are asymptotic to AdS$_4\times S^7$. They are more easily described as solutions in the STU model, a four-dimensional $\cN{=}2$ gauged supergravity with three vector multiplets, which is a consistent truncation both of M-theory on $S^7$, and of the 4d maximal $\cN{=}8$ $SO(8)$ gauged supergravity \cite{Duff:1999gh}. The model contains four Abelian vector fields (one is the graviphoton) corresponding to the $U(1)^4 \subset SO(8)$ isometries of $S^7$.

In 4d $\cN{=}2$ supergravities with $n_V$ vector multiplets, one can use the standard machinery of special geometry \cite{deWit:1984wbb,*Strominger:1990pd,*Andrianopoli:1996cm}. The Lagrangian $\mathscr{L}$ of the theory is completely specified by the prepotential $\cF(X^\Lambda)$, which is a homogeneous holomorphic function of sections $X^\Lambda$, and the vector of Fayet-Iliopoulos (FI) terms $\cG = (g^\Lambda, g_\Lambda)$. The symplectic index $\Lambda = 0, 1, \ldots, n_V$ runs over the graviphoton and the $n_V$ vectors in vector multiplets. The scalars $z^i$ in vector multiplets, with $i=1,\dots, n_V$, parametrize a special K\"ahler manifold $\cM$ and $X^\Lambda$ are sections of a symplectic Hodge vector bundle on $\cM$. The formalism is covariant with respect to symplectic $Sp(2n_V + 2)$ transformations. Indicating as $(A^\Lambda, A_\Lambda)$ the $2n_V+2$ components of a symplectic vector $A$, the scalar product is $\langle A, B\rangle = A_\Lambda B^\Lambda - A^\Lambda B_\Lambda$. One defines the covariantly-holomorphic sections
\be
\cV = e^{\cK(z,\bar z)/2} \mat{ X^\Lambda (z) \\ \cF_\Lambda(z) }
\ee
on $\cM$, where $\cK$ is the K\"ahler potential and $\cF_\Lambda \equiv \partial_\Lambda \cF$. They satisfy $D_\ib \cV \equiv \big( \partial_\ib - \frac12 \partial_\ib \cK\big) \cV = 0$. The K\"ahler potential is then determined by $\langle \cV, \wb\cV \rangle = -i$.

The ansatz for dyonic black holes is of the form
\be
ds^2 = - e^{2U(r)} dt^2 + e^{-2U(r)} \big( dr^2 + V(r)^2 ds^2_{\Sigma_\fg} \big)
\ee
where $\Sigma_\fg$ is a Riemann surface of genus $\fg$, and the scalar fields $z^i$ are assumed to only have radial dependence. We can write the metric on $\Sigma_\fg$ locally as
\be
ds^2_{\Sigma_\fg} = d\theta^2 + f_\kappa^2(\theta)\, d\varphi^2 \;,\quad f_\kappa(\theta) = \left\{ \begin{array}{ll} \sin\theta & \kappa = 1 \\ \theta & \kappa=0 \\ \sinh\theta & \kappa=-1 \end{array} \right.
\ee
where $\kappa = 1$ for $S^2$, $\kappa = 0$ for $T^2$, and $\kappa = -1$ for $\Sigma_\fg$ with $\fg>1$.
The scalar curvature is $2\kappa$ and the volume is
\be
\Vol(\Sigma_\fg) = 2\pi \eta \;, \quad \eta = \left\{ \begin{array}{ll} 2 |\fg-1| &\text{for } \fg \neq 1 \\ 1 &\text{for } \fg = 1 \;. \end{array} \right.
\ee
The magnetic and electric charges of the black hole are
\be
\int_{\Sigma_\fg} \! F^\Lambda = \Vol(\Sigma_\fg) \, p^\Lambda \;,\quad \int_{\Sigma_\fg} \! G_\Lambda = \Vol(\Sigma_\fg) \, q_\Lambda \;,
\ee
where $G_\Lambda = 8 \pi G_\text{N}\, \delta (\mathscr{L} \dvol_4)/\delta F^\Lambda$ and $G_\text{N}$ is the Newton constant. This particular normalization ensures that the BPS equations are independent of $\fg$ (besides a linear constraint). The charges are collected in the vector $\cQ = (p^\Lambda,  q_\Lambda)$. The vector $\cG$ of FI terms controls the gauging and determines the charges of the gravitini under the gauge fields. In a frame with purely electric gauging $g_\Lambda$, the lattice of electro-magnetic charges is
\be
\label{charge quantization}
\eta \, g_\Lambda \, p^\Lambda \in \bZ \;,\qquad \frac{\eta}{4G_\text{N} g_\Lambda} \, q_\Lambda \in \bZ
\ee
not summed over $\Lambda$. It turns out that the BPS equations fix the more stringent condition
\be
\langle \cG, \cQ \rangle = - \kappa \;,
\ee
that we call the linear constraint.

It has been noticed in \cite{Dall'Agata:2010gj} that the BPS equations of gauged supergravity for the near-horizon geometry can be put in the form of ``attractor equations''. One defines the central charge of the black hole $\cZ$ and the superpotential $\cL$:
\bea
\cZ &= \langle \cQ, \cV \rangle = e^{\cK/2} \big( q_\Lambda X^\Lambda - p^\Lambda \cF_\Lambda \big) \\
\cL &= \langle \cG, \cV \rangle \, = e^{\cK/2} \big( g_\Lambda X^\Lambda - g^\Lambda \cF_\Lambda \big) \;.
\eea
The BPS equations for the near-horizon geometry
\be
ds^2_\text{nh} = - \frac{r^2}{R_A^2} \, dt^2 + \frac{R_A^2}{r^2} \, dr^2 + R_S^2 \, ds^2_{\Sigma_\fg}
\ee
with constant scalar fields $z^i$ imply the following two equations \cite{Dall'Agata:2010gj}: $\cZ - i\, R_S^2 \cL = 0$ and $D_j \big( \cZ - i\, R_S^2 \cL \big) = 0$,
where $D_j = \partial_j + \frac12 \partial_j\cK$, besides $\langle \cG, \cQ \rangle = -\kappa$. These equations can be rewritten as
\be
\label{attractor}
\partial_j \frac\cZ\cL = 0 \;,\qquad\quad -i \, \frac\cZ\cL = R_S^2 \;.
\ee
In other words, the scalars $z^i$ at the horizon take a value such that the quantity $-i \cZ/\cL$ has a critical point on $\cM$ and then its value is proportional to the Bekenstein-Hawking black hole entropy.

Notice that a condition to have BHs with smooth horizon is that $-i \cZ/\cL$ be real positive at the critical point. Since the critical-point equations already fix the values of the scalars, this condition becomes a second (non-linear) constraint on the charges. Therefore the domain of allowed electro-magnetic charges has real dimension $2n_V$ (before imposing quantization). There are other inequalities to be satisfied by the charges, for instance to ensure that also $R_A^2$ be positive.

In the case of very special K\"ahler geometry, \ie{} that the prepotential takes the form $\cF = d_{ijk} X^i X^j X^k / X^0$ or symplectic transformations thereof, general solutions to the near-horizon BPS equations as well as full BH solutions have been found in \cite{Halmagyi:2013qoa,*Halmagyi:2014qza,*Katmadas:2014faa}. That analysis guarantees that all near-horizon solutions can be completed into full BH solutions.

Our focus is on the STU model, which has $n_V = 3$ and prepotential
\be
\label{STU prepotential}
\cF = -2i \sqrt{X^0 X^1 X^2 X^3} \;,
\ee
with purely electric gauging $g_\Lambda \equiv g$, $g^\Lambda = 0$. Then the AdS$_4$ vacuum has radius $L^2 = 1/2g^2$. Note that all dyonic BH solutions with electric charges have complex profiles for the scalars, \ie{} the axions are turned on.

{\it The dual field theory.}---%
M-theory on AdS$_4 \times S^7$ has a dual holographic description as a three-dimensional supersymmetric gauge theory, the ABJM theory \cite{Aharony:2008ug}, which provides a non-perturbative definition thereof.
In $\cN{=}2$ notation, the ABJM theory is a $U(N)_1 \times U(N)_{-1}$ Chern-Simons theory (the subscripts are the levels) with bi-fundamental chiral multiplets $A_i$ and $B_j$, $i,j = 1,2$,  transforming in the $(N,\overline N)$ and $(\overline N, N)$ representations of the gauge group, respectively, and with superpotential $W = \varepsilon^{ik} \varepsilon^{jl} \Tr A_i B_j A_k B_l$. The theory has $\cN{=}8$ superconformal symmetry and $SO(8)$ R-symmetry. The identification between gravitational and QFT parameters is
\be
\label{formula for N}
\frac{L^2}{G_\text{N}} = \frac1{2g^2 G_\text{N}} = \frac{2\sqrt2}3 N^{3/2} \;.
\ee

The ``topologically twisted index'' of an $\cN{=}2$ three-dimensional theory is its supersymmetric Euclidean partition function on $S^2 \times S^1$ with a topological twist on $S^2$ \cite{Benini:2015noa}. Its higher-genus generalization, namely the twisted partition function on $\Sigma_\fg \times S^1$, has been constructed as well \cite{Benini:2016hjo,Closset:2016arn}. They depend on a set of integer magnetic fluxes $\fp_a$ and complex fugacities $y_a$, along the Cartan generators of the flavor symmetry group.

In the present case, to make the enhanced symmetry more manifest, we introduce an index $a=1,2,3,4$ that simultaneously runs over the four ABJM chiral fields and the four Abelian symmetries $U(1)^4 \subset SO(8)$. This is done by introducing a basis of four R-symmetries $R_a$, each acting with charge $2$ on one of the chiral fields and zero on the others. Then the magnetic fluxes identify a $U(1)$ subgroup of $SO(8)$ used to twist, and are required by supersymmetry to satisfy $\sum_a \fp_a = 2g-2$. The complex fugacities $y_a = \exp iu_a$ must satisfy $\prod_a y_a =1$ ($\sum_a u_a \in 2\pi \bZ$) and encode background values for the flavor symmetries. Writing $u_a = \Delta_a + i \beta \sigma_a$ (where $\beta$ is the length of $S^1$), we can identify $\Delta_a$ with flavor flat connections and $\sigma_a$ with real masses.

The Hamiltonian definition of the index is \cite{Benini:2015noa}
\be
\label{def twisted index}
Z(u_a,\fp_a) = \Tr\, (-1)^F \, e^{i \sum_{a=1}^3 \Delta_a J_a} \, e^{-\beta H} \;,
\ee
where $J_a = \frac12(R_a -R_4)$ are the three independent flavor symmetries and $H$ is the twisted Hamiltonian on $S^2$, explicitly dependent on the magnetic charges $\fp_a$ and the real masses $\sigma_a$. Due to the supersymmetry algebra $Q^2= H -\sum_{a=1}^3  \sigma_a J_a$, the index $Z(u_a,\fp_a)$ is a meromorphic function of $y_a$. For simplicity, we will keep the dependence on $\fp_a$ implicit and use the shorthand notation $\sigma J =\sum_{a=1}^3 \sigma_a J_a$. We stress that, in general, (\ref{def twisted index}) is well-defined only for complex $u_a$ while the index for $\sigma_a=0$ is defined by analytic continuation. We would like to see how we can extract the BH entropies from $Z$.

{\it Statistical interpretation.}---%
The partition function $Z(u)$  describes a supersymmetric ensemble which is canonical with respect to the magnetic charges (\ie{} all states have the same, fixed, magnetic charges) but grand canonical with respect to the electric charges (\ie{} it is a sum over all electric charge sectors, with fixed chemical potentials $u_a$). A similar viewpoint in BH physics is advocated in \cite{Hawking:1995ap}. We can decompose $Z$ as a sum over sectors with fixed charges $\fq_a$ under $R_a/2$ (then the lattice of charges is such that both $J_a, R_a \in \bZ$, up to a possible zero-point shift in the vacuum):
\be
Z(u) = \sum\nolimits_\fq \, e^{i \sum_{a=1}^3 u_a (\fq_a - \fq_4)} \, Z_\fq \;.
\ee
We would like to identify $S_\fq \equiv \re \log Z_\fq$ with the leading entropy of a BH of fixed electric charges $\fq_a$. We take the real part to remove the effect of a possible overall sign. An important assumption is that $(-1)^F$ in the trace (\ref{def twisted index}) does not cause dangerous cancelations at leading order.

We can Fourier transform the previous expression with respect to the three independent $\Delta_a$ to obtain
\be
\label{Fourier}
\sum\nolimits_{\fq_4}' Z_\fq = \int \frac{ d^3\Delta_a}{(2\pi)^3} \, e^{-i \sum_{b=1}^3 \Delta_b (\fq_b - \fq_4)} \, Z(u) \;,
\ee
where prime means that the sum is taken at fixed integer $\fq_a - \fq_4$. As we will see, for supergravity BHs both the electric charges $\fq_a$ and $\log Z$ are of order $N^{3/2}$, therefore the previous expression can be evaluated at large $N$ using a saddle point approximation:
\be
\label{saddle}
\sum\nolimits_{\fq_4}' Z_\fq = \exp \Big[ \log Z(\hat u) - i \sum\nolimits_{a=1}^3 \hat u_a (\fq_a - \fq_4) \Big]
\ee
at leading order, where $\hat u_a$ is a solution for $u_a$ to
\be
\label{saddleextremization}
\parfrac{}{u_a} \Big[ \log Z(u) - i \sum\nolimits_{b=1}^3 u_b (\fq_b - \fq_4)\Big] = 0
\ee
with $a = 1,2,3$. This saddle point in general gives complex values for $\hat u_a$. The sum on the LHS of (\ref{saddle}) will also be dominated by a specific value of $\fq_4$, corresponding to the electric R-charge of the black hole. For that value:
\be
\label{entropy}
S_\fq = \re \Big[ \log Z(\hat u) - i \sum\nolimits_{a=1}^3 \hat u_a(\fq_a - \fq_4) \Big] \;.
\ee
We can restore the permutation symmetry between the charges, part of the Weyl group of $SO(8)$, by introducing
\be
\cI(u) \equiv \log Z(u) - i \sum\nolimits_{a=1}^4 u_a \fq_a \;.
\ee
Eqn. (\ref{saddleextremization}) is equivalent to extremization of $\cI$ and the entropy is given by  $S_\fq = \re \cI (\hat u)$.

This argument does not determine the R-charge of the BH, essentially because the index $Z(u)$ lacks a chemical potential for it. However from the attractor equations (\ref{attractor}) it follows that, for given magnetic charges $\fp_a$ and flavor electric charges $\fq_a - \fq_4$, there is at most one value of $\fq_4$ leading to a large smooth BH. Our argument then gives an unambiguous prediction for the leading entropy of that BH.

{\it RG flow interpretation.}---%
We can extract more information from the index if we interpret the BH as an holographic RG flow. The near-horizon geometry of BPS black holes contains an AdS$_2$ factor permeated by constant electric flux, where the super-isometry algebra is enhanced to $\su(1,1|1)$. Thus we can think of the BH solution as a holographic RG flow from the 3d theory on $S^2$ to an ensemble of $\su(1,1|1)$-invariant states in a 1d system. The bosonic subalgebra is $\mathfrak{sl}(2,\bR) \times \mathfrak{u}(1)_c$ where the second factor is the IR superconformal R-symmetry, which is some linear combination of $U(1)^4 \subset SO(8)$. In the near-horizon canonical ensemble this implies that all BH states have zero $U(1)_c$ charge (by an argument similar to that in \cite{Sen:2009vz,*Dabholkar:2010rm}). We will assume that there are no other contributions outside the horizon.

The asymptotic behavior of electrically charged BH solutions with axions turned on suggests that the dual ABJM theory is also deformed by real masses $\sigma_a$. In general, they lift a possible vacuum degeneracy of the Hamiltonian $H$. The presence of AdS$_2$ with constant electric flux, though, indicates that there should be a large vacuum degeneracy for a modified Hamiltonian $H_\text{nh}$ in which the energy of states gets an extra contribution linear in the charge: $H_\text{nh}(\sigma) = H(\sigma) - \sigma$. From the supersymmetry algebra $Q^2=H_\text{nh}$ we conclude that $H_\text{nh} \geq 0$, and the index gets contribution only from its ground states. We can rewrite the index in (\ref{def twisted index}) as
\be
\label{index in BH form}
Z(\Delta,\sigma)  = \Tr^\prime e^{i\pi R_\text{trial}(\Delta)} \, e^{-\beta \sigma J} \;,
\ee
where $\Tr^\prime =\Tr_{H_\text{nh}=0}$. We introduced a trial R-current $R_\text{trial}(\Delta) \equiv R_0 + \Delta \, J/\pi$ that parametrizes the mixing of the R-symmetry with the flavor symmetries, with $R_0$ a reference R-symmetry such that $e^{i\pi R_0} = (-1)^F$.

We want to argue, generalizing \cite{Benini:2015eyy}, that the superconformal R-symmetry $R_c$ of the Hamiltonian $H_\text{nh}$ can be found by extremizing  $Z(\Delta,\sigma)$ for fixed values of $\sigma_a$. Let  $\hat \Delta_a$ be the value such that $R_\text{trial}(\hat\Delta) = R_c$. One computes $\partial \log Z/ \partial \Delta_a \big|_{\hat \Delta_a} = i \langle J_a e^{-\beta\sigma J} \rangle / \langle e^{-\beta\sigma J} \rangle$, using that at zero temperature the density matrix is uniformly distributed over the ground states of $H_\text{nh}$, and that $R_c=0$ in those states as argued above. The expression on the right is imaginary, implying that $\hat\Delta_a$ are determined by extremizing the index with respect to $\Delta_a$ at fixed $\sigma_a$:
\be
\parfrac{\re\log Z(\Delta,\sigma)}{\Delta_a} \Big|_{\hat\Delta} = 0 \;.
\ee
This is the generalization of the $\cI$-extremization principle proposed in \cite{Benini:2015eyy}. Assuming the large $N$ factorization $\langle J e^{-\beta \sigma J} \rangle = \langle J \rangle \langle e^{-\beta \sigma J} \rangle$, we also have
\be
\parfrac{\im \log Z(\Delta,\sigma)}{\Delta_a} \Big|_{\hat\Delta} = i \langle J_a \rangle \equiv i (\fq_a - \fq_4) \;,
\ee
where $\langle J_a \rangle$ is the charge of the vacuum density matrix. This determines the relation between the flavor charges $\fq_a-\fq_4$ and $\sigma_a$. Since $Z(\Delta,\sigma)$ is a holomorphic function of $u_a = \Delta_a + i\beta\sigma_a$, we can summarize the result in the complex equation
\be
\label{complexextremization}
\parfrac{\log Z(u)}{u_a} \Big|_{\hat u} = i (\fq_a-\fq_4) \;,
\ee
which determines both $\hat\Delta_a$ and $\sigma_a$ as functions of $\fq_a$.

From eqn. (\ref{index in BH form}), at the critical point
\be
Z(\hat \Delta, \sigma) = e^{-\beta\sigma \langle J \rangle} \Tr' 1 = e^{-\sum_{a=1}^4 \beta \sigma_a \fq_a} \, e^{S_\fq} \;.
\ee
The real part of the logarithm of this expression reproduces the result of the statistical argument, namely
\be
\label{entropy in brackets}
S_\fq = \re \Big[ \log Z(\hat u) - i \sum\nolimits_{a=1}^4 \hat u_a \fq_a \Big] \;.
\ee
An advantage of this derivation is that we can argue, at least at leading order, that $e^{S_\fq}$ is the number of ground states, without dangerous signs that could cause cancelations.

We can also write the entropy in a slightly different form and make a conjecture for the value of the fourth charge. Since $\hat u$ only depends on the differences \mbox{$\fq_a - \fq_4$} and $\sum_a u_a \in 2\pi \bZ$, we can always shift the integer charges $\fq_a$ and write the entropy in the permutationally symmetric and holomorphic form
\be
\label{entropy equation}
S_\fq = \log Z(\hat u) - i \sum\nolimits_{a=1}^4 \hat u_a \fq_a = \cI(\hat u) \;,
\ee
up to $\cO(N^0)$ terms which are invisible in the large $N$ limit.
The determination of the logarithm is such that $\log Z$ is real for $\sigma_a=0$ and extended by continuity. The requirement that (\ref{entropy equation}) be real positive fixes the fourth charge. Interestingly, this is precisely the constraint (\ref{attractor}) that comes from supergravity.

{\it Explicit match for ABJM}.---%
The large $N$ expression for the index of ABJM was found in \cite{Benini:2015eyy, Benini:2016hjo} for the case of real $u_a$, and we can extend it to the complex plane using holomorphy:
\be
\label{ABJMindex}
\log Z = \frac{N^{3/2}}3 \sqrt{2 u_1 u_2 u_3 u_4} \, \sum\nolimits_{a=1}^4 \frac{\fp_a}{u_a} \;.
\ee
This is valid  for $\sum\nolimits_a u_a = 2\pi$  and  $0 < \re u_a < 2\pi$.
The $\cI$-extremization principle (\ref{complexextremization}) is equivalent to the extremization of
\be
\label{I-ext field theory}
\cI_\text{QFT} =  \sum\nolimits_{a=1}^4 \bigg( \frac{N^{3/2}}3 \sqrt{2 u_1 u_2 u_3 u_4} \,  \frac{\fp_a}{u_a} - i  \fq_a  u_a \bigg) \;.
\ee
Then the entropy is given by $S_\fq = \cI_\text{QFT}(\hat u)$, with the constraint on the charges that $\cI_\text{QFT}(\hat u)$ be positive.

In supergravity, the BH entropy is determined by
\be
S_\text{BH} = \frac{\text{Area}}{4G_\text{N}} = - i \frac{\cZ}{\cL} \, \frac{2\pi \eta}{4G_\text{N}} \equiv \cI_\text{SUGRA}
\ee
using (\ref{STU prepotential}), and $\cI_\text{SUGRA}$ should be extremized with respect to $X^\Lambda$. We can identify the index $\Lambda = \{0,1,2,3\}$ with $a = \{1,2,3,4\}$, as well as $2\pi X^a / \sum_b X^b = u_a$ since they have the same domain and constraint:
\be
\cI_\text{SUGRA} = \frac{\eta}{4gG_\text{N}} \sum\nolimits_{a=1}^4 \bigg( \sqrt{u_1 u_2 u_3 u_4} \, \frac{p^a}{u_a} - i q_a u_a \bigg) \;.
\ee
Identifying the integers in (\ref{charge quantization}) with the charges $\fp_a$, $\fq_a$, respectively, and using (\ref{formula for N}) we obtain a perfect match $\cI_\text{QFT} = \cI_\text{SUGRA}$. The field theory extremization principle corresponds to the supergravity attractor mechanism: they lead to the same entropy and non-linear constraint on the charges.

\begin{acknowledgements}
We thank J. de Boer, A. Gnecchi, N. Halmagyi and S. Murthy for instructive clarifications. FB is supported by the MIUR-SIR grant RBSI1471GJ. AZ is supported by the MIUR-FIRB grant RBFR10QS5J.
\end{acknowledgements}

\bibliography{BHEntropy_PRL}

\end{document}